\documentclass[a4paper,12pt]{article}
\usepackage{cite}
\newcommand{\be}{\begin{equation}}
\newcommand{\ee}{\end{equation}}
\newcommand{\bea}{\begin{eqnarray}}

\newcommand{\eea}{\end{eqnarray}}

\begin{document}

\begin{titlepage}
\begin{flushright}

\end{flushright}
\begin{centering}
\vspace{.3in}
{\Large{\bf Self-gravitational corrections to the Cardy-Verlinde formula of charged  BTZ black hole}}
\\

\vspace{.5in} {\bf Farhad Darabi$^{1}$,  Mubasher Jamil$^{2}$ and Mohammad
Reza Setare$^{3}$} \\
\vspace{.3in} $^{1}$\,  Department of Physics, Azarbaijan University of Tarbiat Moallem, Tabriz, Iran\\
f.darabi@azaruniv.edu\\ Corresponding author\\
\vspace{0.4in} $^{2}$\, Center for Advanced Mathematics and Physics,\\ 
National University of Sciences and Technology, H-12, Islamabad, Pakistan\\
mjamil@camp.nust.edu.pk\\
\vspace{0.4in} $^{3}$\, Department of Science, Payame Noor University, Bijar,
Iran\\
rezakord@ipm.ir\\
\vspace{0.4in}
\end{centering}

\vspace{0.6in}
\begin{abstract}
The entropy of the charged BTZ black hole horizon is
described by the Cardy-Verlinde formula. We then compute the self-gravitational corrections to
the Cardy-Verlinde formula of the charged  BTZ black hole in the context of Keski-Vakkuri, Kraus and Wilczek (KKW)
analysis. The self-gravitational
corrections to the entropy as given by the Cardy-Verlinde formula
are found to be positive. This result provides evidence in  support of the claim that the holographic
bound is not universal in the framework of two-dimensional gravity models.
\end{abstract}
\end{titlepage}

\newpage

\baselineskip=18pt
\section*{Introduction}
It is well known that ($2+1$)-dimensional gravity
has a black hole solution which was first obtained in 1992 by Ba$\tilde{n}$ados, Teitelboim and Zanelli (BTZ)
\cite{banados1, banados2}. This black hole solution was first described by 
its mass $M$ and its angular momentum (spin) $J$ over a locally
anti-de-Sitter space and thus it would differ from Schwarzschild and
Kerr solutions in that it was asymptotically anti-de-Sitter instead
of flat spacetime. However, it turned out later that \cite{Clement1, Clement2,
Clement3, Clement4, Clement5, Clement6}
if one includes electric charge, the solution reported in the original articles
\cite{banados1, banados2} only applies when the angular momentum vanishes.
\par
We know that Hawking effect as a quantum process \cite{hawking1} is usually studied using a fixed background during the emission process. There is another idea of Keski-Vakkuri, Kraus and Wilczek (KKW)
\cite{KKW1, KKW2, KKW3, KKW4} in which the black hole background is considered as dynamical by treating the Hawking radiation as
a tunnelling process. The energy conservation is the key to this description. The total (ADM) mass is kept fixed
while the mass of the black hole under consideration decreases due to the emitted radiation. The effect of
this modification gives rise to additional terms in the formulae concerning the known results for
black holes \cite{correction1, correction2, correction3, correction4, correctio5,
correctio6} such that a nonthermal partner to the thermal spectrum of the Hawking radiation shows up.
\par
Holography is believed to be one of the fundamental principles of the true
quantum theory of gravity \cite{HOL1, HOL2, HOL3, HOL4, HOL5}. An explicitly
calculable example of holography is the much studied anti-de
Sitter (AdS)/Conformal Field Theory (CFT) correspondence. More
recently, it has been proposed in a manner analogous
with the AdS$_{d}$/CFT$_{d-1}$ correspondence, that quantum gravity in a
de Sitter (dS) space is dual to a certain
Euclidean  CFT living on a spacelike boundary of the
dS space~\cite{Strom1, Strom2} (see also earlier works
\cite{mu1, mu2, mu3, mu4}). Following this proposal, some
investigations on the de Sitter space have been carried out recently
\cite{Mazu1, Mazu2, Mazu3, Mazu4, Mazu5, Mazu6, Mazu7, Mazu8, Mazu9,  Mazu10, Mazu11,
Mazu12, Mazu13, Mazu14, Mazu15, Mazu16, Mazu17, Mazu18, Mazu19,  Mazu20, Mazu21, Mazu22, Mazu23, Mazu24, Mazu25, Mazu26, Mazu27}.
\par
The Cardy-Verlinde formula recently proposed by  Verlinde \cite{Verl}, relates the entropy of a  certain CFT to
its total energy and Casimir energy in arbitrary dimensions. In the spirit of AdS$_{d}$/CFT$_{d-1}$ and
dS$_{d}$/CFT$_{d-1}$ correspondences, this formula has been shown to hold exactly for the cases of topological dS,
Schwarzschild-dS, Reissner-Nordstr\"om-dS, Kerr-dS and Kerr-Newman-dS black holes.
\par
Recently, much interest has been taken  in computing the quantum corrections to the
Bekenstein-Hawking entropy $S_{BH}$ \cite{med,muk,lid}. In a recent work Carlip \cite{carlip} has
deduced the leading order quantum correction to the classical Cardy formula. The Cardy formula
follows from a saddle-point approximation of the partition function for a two-dimensional CFT.
This leads to the theory's density of states which is related to the partition function
by way of a Fourier transform.  In \cite{med}, Medved has employed Carlip's formulation
to the case of a generic model of two-dimensional gravity with coupling to a dilaton field. 
In \cite{SV}, Setare (one of the authors of the present paper) and Vagenas have obtained the self-gravitational corrections to
the Cardy-Verlinde formula of the  Ach\'ucarro-Ortiz black hole in the context of Keski-Vakkuri, Kraus and Wilczek (KKW) analysis.
Motivated by this activity, in the present paper, we study the semi-classical gravitational corrections to the Cardy-Verlinde
formula of the charged BTZ black hole due to the effect of self-gravitation.
\par
The remainder of this paper is organized as follows. In Sections 1 and 2, we make a short review of the charged  BTZ black hole and present expressions for its mass, temperature, electric potential, area and entropy \cite{S-J} \footnote{It is to be noted that we will set $J=0$ in all results in \cite{S-J}
which we shall use in the present paper.}.
In Section 3, we compute the self-gravitational
corrections to the entropy of the charged  BTZ black hole which is described by the Cardy-Verlinde formula.

\section{The charged  BTZ black hole }
The BTZ black hole solutions \cite{banados1}, \cite{Martinez:1999qi} in
$(2+1)$ spacetime dimensions with vanishing angular momentum are derived from a three dimensional
theory of gravity 
\be 
\label{ac1}
I=\frac{1}{16 \pi G}\int dx^{3} \sqrt{-g}\, 
({\cal R}+2\Lambda),
\ee
where ${\cal R}$ is the Ricci scalar, $G$ is the three dimensional Newton
constant and $\Lambda=\frac{1}{l^2}>0$ is the cosmological constant.
Often in the literature units are chosen such that $G$ is
dimensionless, $8\pi G=1$, here we use such units..
\par\noindent
The corresponding line element in Schwarzschild coordinates is 
\be
\label{metric}
ds^2 =- f(r)dt^2 + f^{-1}dr^{2}¥+r^2 d\theta
^2, 
\ee 
with metric function:
\be 
\label{metric2}
f(r)=\left(-M+\frac{r^2}{l^2} \right),
\ee 
where $M$ is the Arnowitt-Deser-Misner (ADM) mass,
and $-\infty<t<+\infty$, $0\leq r<+\infty$,
 $0\leq \theta <2\pi$.
The horizon of the line element (\ref{metric}) corresponding to the positive mass spectrum is given by
\be
\label{horizon1}
r^{2}=M{l^2}.  
\ee 
In addition to the BTZ solutions described above, there are solutions
following from the action \cite{Martinez:1999qi,Achucarro:1993fd}
\be 
\label{ac2}
I=\frac{1}{2\pi }\int dx^{3} \sqrt{-g}\,(
({\cal R}+2\Lambda-\frac{\pi}{2} F_{\mu\nu}F^{\mu\nu}) . 
\ee 
The Einstein equations are given by 
\be 
\label{ein}
G_{\mu\nu}-\Lambda
g_{\mu\nu}=\pi T_{\mu\nu}, 
\ee 
where $T_{\mu\nu}$ is the
energy-momentum tensor of the electromagnetic field: 
\be 
\label{tmn}
T_{\mu\nu}=F_{\mu\rho}F_{\nu\sigma}g^{\rho\sigma}-\frac{1}{4}g_{\mu\nu}F^2.
\ee
 Electrically charged black hole solutions of the equations (\ref{ein})
 takes the
form (\ref{metric}), but with 
\be 
\label{charged}
f(r)=-M+\frac{r^2}{l^2} -\frac{\pi}{2} Q^2 \ln r.
\hspace{0.5cm} 
\ee 
The $U(1)$ Maxwell field is also given by
\be
\label{maxw} 
F_{tr}=\frac{Q}{r},\label{metric3}
\ee
where $Q$ is the electric charge. 
The line element of the electrically charged  BTZ black hole with
the function $f(r)$ given by Eq.(\ref{charged}) has two {\it formal} horizons
\be
r_{\pm}=\pm\frac{1}{2}\sqrt{2Ml^2-Q\Phi_{\pm}l^2\pm l\sqrt{4M^2l^2-4Ml^2Q\Phi_{\pm}
+Q^2\Phi_{\pm}^2l^2}},\label{r''}
\ee
where $r_+$, $r_-$ are the outer and inner horizons, respectively, and
\be 
\Phi_{\pm}=\left.\frac{\partial
M}{\partial Q}\right\vert_{r=r_{\pm}}= -\pi Q \ln r_{\pm}.
\ee 

Horizons of the charged BTZ metric are roots of the lapse function $f$.
Depending on these roots there are three cases of the charged BTZ black
hole \cite{akbar}: 1) two distinct horizons $r_{\pm}$ exist where plus
correspond to the event horizon while minus gives the Cauchy horizon
; 2) black hole in case of two repeated real roots
gives a single horizon (extreme case); and 3) the case when no real
root exists thus no horizon exists (naked singularity).

We shall assume the first case in this paper. The black hole mass
is given by  
\be
M=\frac{r_+^2}{l^2}-\frac{\pi}{2}Q^2\ln r.
\ee 
The Hawking temperature $T_{H}$ of the black hole horizon
is 
\be 
T_{H}=\left.\frac{df}{dr}\right\vert_{r=r_+}=\frac{1}{4\pi}
\Big(
\frac{2r_+}{l^2}-\frac{\pi}{2}\frac{Q^2}{r_+}
\Big). 
\ee 
The entropy of the charged  BTZ black hole takes
the form 
\be 
\label{14} 
S_{BH}=4\pi r_+. 
\ee 
Also the electric
potential of the black hole is 
\be 
\Phi_+=\left.\frac{\partial
M}{\partial Q}\right\vert_{r=r_+}= -\pi Q \ln r_+.
\ee 
The generalized Cardy formula (hereafter named Cardy-Verlinde formula)
is given by 
\be 
\label{16} 
S_{CFT}=\frac{2\pi
R}{\sqrt{ab}}\sqrt{E_C(2E-E_C)}, 
\ee 
where $R$ is the radius of the sphere, $E$ is the total energy
and $E_C$ is the Casimir energy. The definition of the Casimir
energy is derived by the violation of the Euler relation 
\be
\label{17} 
E_C=n(E+PV-TS-\Phi_+ Q), 
\ee 
where the pressure
of the CFT is defined as $P=E/nV$. The total energy may be written
as the sum of two terms 
\be 
\label{18} 
E=E_E+\frac{1}{2}E_C, 
\ee
where $E_E$ is the purely extensive part of the total energy $E$.
The Casimir energy $E_C$ as well as the purely extensive part of
energy $E_E$ expressed in terms of the radius $R$ and the entropy
$S$ are written as 
\be 
\label{cas}
E_C=\frac{b}{2\pi R}, 
\ee 
\be
\label{ex} 
E_E= \frac{a}{4\pi R}S^2.
\ee

\section{Entropy of charged  BTZ black hole in Cardy-
Verlinde formula}
The Casimir energy $E_C$, defined
as Eq.(\ref{17}), and $n=1$ in this case, is found to be 
\be
\label{21} 
E_C=\frac{1}{2}\Big( \pi Q^2 \Big). 
\ee
Additionally, it is obvious that 
\be 
\label{20} 
2E-E_C=\frac{2r^2_+}{l^2}- \pi Q^2\Big( \ln r_++\frac{1}{2} \Big). 
\ee 
The purely extensive part of the total energy $E_E$ by substituting
Eq.(\ref{20}) in Eq.(\ref{18}), is given as
\be 
\label{23} 
E_E=\frac{r^2_+}{l^2} - \frac{\pi}{2} Q^2\Big( \ln r_++\frac{1}{2} \Big),
\ee 
whilst by substituting Eq.(\ref{14}) in Eq.(\ref{ex}), it takes
the form 
\be 
\label{24} 
E_E=\frac{4\pi a}{R}r_+^2. 
\ee 
Making use of expression (\ref{cas}), Casimir energy $E_C$ can also be written as
\be 
\label{25} 
E_C=\frac{b}{2\pi R}. 
\ee 
At this point it is useful to evaluate the radius $R$. By equating the right hand sides of (\ref{21}) and (\ref{25}), the radius is written as 
\be
R=\frac{b}{\pi^2  Q^2 }, \label{r} 
\ee 
while by equating the right hand sides (\ref{23}) and (\ref{24}) it can also
be written as 
\be 
R=\frac{4\pi ar_+^2l^2}{r_+^2-\frac{\pi}{2}
Q^2l^2\Big( \ln r_++\frac{1}{2} \Big)}. \label{r'} 
\ee 
By multiplication of both sides in Eqs.(\ref{r}), (\ref{r'}) and taking its
square root, the radius expressed in terms of the arbitrary positive coefficients $a$ and $b$ is obtained
\be 
\label{28} 
R=\frac{2r_+l\sqrt{ab}}{\sqrt{
\pi Q^2 r_+^2-\frac{\pi^2}{2} Q^4l^2\Big( \ln
r_++\frac{1}{2} \Big)}}. 
\ee 
Finally, we substitute expressions
(\ref{21}), (\ref{20}) and (\ref{28}) which were derived in the
context of thermodynamics of the charged  BTZ black hole, in
the Cardy- Verlinde formula (\ref{16}) which in turn was derived in
the context of CFT, and we get 
\be 
S_{CFT}=S_{BH}. 
\ee 
It has been proven that the entropy of the charged  BTZ black hole can be expressed in the form of Cardy-Verlinde formula.
\section{Self-gravitational corrections to Cardy-Verlinde formula}
In this section we compute the self-gravitational corrections to the entropy of the charged  BTZ black hole (\ref{14}) described by the Cardy-Verlinde
formula
\be
\label{C-V}
S_{CFT}=\frac{2\pi
R}{\sqrt{ab}}\sqrt{E_{C}\left(2E-E_{C}\right)}
\hspace{1ex}.
\ee
The total energy $E$ may be written as the sum of two terms
\be
E(S, V)=E_{E}(S, V)+\frac{1}{2}E_{C}(S,V),
\label{ext}
\ee
where $E_{E}$ is the purely extensive part of the total energy $E$ and $E_{C}$ is the Casimir energy.
\par\noindent
The Casimir energy is derived by the violation of the Euler relation
\be
E_{C}=2 E-T_{H}S_{BH} -Q\Phi_+,
\label{euler} 
\ee 
which now  will be modified due to the self-gravitation effect as 
\be 
\tilde{E}_{C}=2E-T_{bh}S_{bh}-Q_{bh}\Phi_{bh}
\label{euler1}
\hspace{1ex}.
\ee
Using the KKW analysis \cite{correction5} and after long calculations it is straightforward to show that 
\be
T_{bh}S_{bh}=T_{H}S_{BH}-\omega \frac{M}{\left(M-\frac{1}{2}Q\Phi_+ +\sqrt{M^2 -M Q\Phi_+ +\frac{Q^2\Phi_+^2}{4}}\right)},
\label{corts}
\ee
where $\omega$ is the emitted shell of energy radiated outwards the black hole horizon.

At this point it is necessary to stress that we shall consider no self-gravitational
corrections to the total energy $E$ of the system under study as well as to the radius (\ref{28}).\\
It is also obvious that no self-gravitational corrections should exist for the fourth term in (\ref{euler1}), namely the electric term $Q \Phi_+$. This
is because, the emitted (neutral) shell of energy can neither carry away the electric charge $Q$ nor alter the electric potential $\Phi_+$ of the black hole. Therefore,
we have 
\be
Q_{bh}\Phi_{bh} =Q\Phi_{+}.
\ee
The modified Casimir energy is then obtained
\bea
\label{euler2}
\tilde{E}_{C}&=& 
\frac{1}{2}\Big( \pi Q^2 \Big)+\omega \frac{M}{\left(M-\frac{1}{2}Q\Phi_+
+\sqrt{M^2 -M Q\Phi_+ +\frac{Q^2\Phi_+^2}{4}}\right)}\\ \nonumber
&=&E_{C}+2\epsilon_{2} M\,\omega,
\eea
$\epsilon_{2}$ being a parameter given by
\bea
\epsilon_{2}& =& 
\frac{1}{2\left(M-\frac{1}{2}Q\Phi_+ +\sqrt{M^2 -M Q\Phi_+ +\frac{Q^2\Phi_+^2}{4}}\right)}\\
\nonumber
&=&\frac{l^2}{4 r_{+}^{2}}
\hspace{1ex},
\eea
where use has been made of (\ref{r''}) to obtain the second equality. Additionally, it is evident that the modified quantity
$2E-\tilde{E}_{C}$ is given by 
\bea 
\label{core-ec}
2E-\tilde{E}_{C}&=&\\ \nonumber
& &2\frac{r^2_+}{l^2}- \pi Q^2\Big( \ln r_++\frac{1}{2} \Big)-\omega \frac{M}{\left(M-\frac{1}{2}Q\Phi_+
+\sqrt{M^2 -M Q\Phi_+ +\frac{Q^2\Phi_+^2}{4}}\right)}\\ \nonumber
&=&2E-E_{C}-2\epsilon_{2} M\,\omega
\hspace{1ex}.
\eea
Apart from the Casimir energy, the purely extensive part of the total energy $E_{E}$ will also be modified
due to the effect of self-gravitation. Thus, it takes the form
\bea
\label{corexten2}
\tilde{E}_{E}&=& 
\frac{r^2_+}{l^2} - \frac{\pi}{2} Q^2\Big( \ln r_++\frac{1}{2} \Big)-\omega \frac{M}{2\left(M-\frac{1}{2}Q\Phi_+ +\sqrt{M^2 -M Q\Phi_+ +\frac{Q^2\Phi_+^2}{4}}\right)}\\
\nonumber
&=&E_{E}-\epsilon_{2} M\,\omega,
\eea
whilst it can also be written as \cite{e-m}
\bea
\label{corexten3}
\tilde{E}_{E}&=&\frac{a}{4\pi R}S_{bh}^{2}\\ \nonumber
&=&\frac{4\pi a}{R}r^{2}_{out}\\ \nonumber
&=&\frac{4\pi a}{R}r^{2}_{+}(1-2\epsilon_{1}\omega)
\hspace{1ex}.
\eea
\par\noindent
Substituting the expressions (\ref{28}), (\ref{euler2}) and (\ref{core-ec})
in the Cardy-Verlinde formula (\ref{C-V}),the self-gravitational corrections result in
\bea
S_{CFT}&=&\frac{2\pi}{\sqrt{ab}}\frac{2r_+l\sqrt{ab}}{\sqrt{
\pi Q^2 r_+^2-\frac{\pi^2}{2} Q^4l^2\Big( \ln
r_++\frac{1}{2} \Big)}}\\ \nonumber
&\times&\sqrt{\left(\frac{1}{2}\pi Q^2 +2\epsilon_{2} M\,\omega\right)
\left(2\frac{r^2_+}{l^2}- \pi Q^2\Big( \ln r_++\frac{1}{2} \Big)-2\epsilon_{2} M\,\omega\right)},
\eea
and consequently, the first-order self-gravitationally corrected
Cardy-Verlinde formula of the charged  BTZ black hole takes the form
\be
S_{CFT}=S_{BH}\sqrt{1+\epsilon_{3}\omega} \label{r'''}
\hspace{1ex}.
\ee
where
\bea
\epsilon_{3}&=&4M
\left(\frac{r_{+}^{2}}{\pi Q^2 r_+^2}-\frac{l^2}{4r^{2}_{+}-\pi Q^2l^2(2\ln
r_+ +1)}\right)\epsilon_{2}.
\eea
It is easily seen that this correction is positive, namely one may
show that the modified term $\tilde{E}_{C}(2E-\tilde{E}_{C})$ in $S_{CFT}$ is
larger than the original one ${E}_{C}(2E-{E}_{C})$. To this end, we expand the modified term $\tilde{E}_{C}(2E-\tilde{E}_{C})$
to the first order in terms of $\omega$ as
\bea
\tilde{E}_{C}(2E-\tilde{E}_{C})&=&\Big(\frac{1}{2}\pi Q^2 +2\epsilon_{2} M\,\omega\Big)
\Big(2\frac{r^2_+}{l^2}- \pi Q^2\Big( \ln r_++\frac{1}{2} \Big)-2\epsilon_{2} M\,\omega\Big)\\ \nonumber
&=&{E}_{C}(2E-{E}_{C})+2\epsilon_{2} M\,\omega\Big((2E-E_{C})-E_{C}\Big).
\eea
Now, according to (\ref{18}), $E>E_{C}$ and so $2E-2E_{C}>0$. Therefore, with $\epsilon_{2}>0$ we have $\tilde{E}_{C}(2E-\tilde{E}_{C})>{E}_{C}(2E-{E}_{C})$.
In fact, the physical reason why the first-order self-gravitational correction to the Cardy-Verlinde formula of the charged BTZ black hole is positive lies
in the requirement $\epsilon_3>0$. This is because in the contrary case, it may yield the right hand side of (\ref{r'''}) to be imaginary whereas the LHS must be a real and positive quantity.

It should be pointed out that in the context of KKW analysis the self-gravitational corrections
to the entropy as described by the Cardy-Verlinde formula ($S_{CFT}$) of the
charged  BTZ black hole are different from the ones to
the corresponding Bekenstein-Hawking entropy ($S_{BH}$).
This is expected since in order to evaluate the corrections to the entropy as described
by the Cardy-Verlinde formula ($S_{CFT}$), we have taken into account not only corrections to the
Bekenstein-Hawking entropy but also to all quantities appearing in the Cardy-Verlinde formula except for the charge $(Q)$ and electric potential
$(\Phi_+)$ which have no such corrections.
Furthermore, the entropy of the charged  BTZ black hole ($S_{CFT}$)
described in the context of KKW analysis by the semiclassically corrected
Cardy-Verlinde formula violates the holographic bound \cite{HOL1,HOL2, HOL3 }, i.e.
\be
S_{CFT}>S_{BH}>S_{bh}
\hspace{1ex}.
\ee


\section{Conclusion}

In this paper we have evaluated the
semi-classical corrections to the entropy of a charged rotating BTZ
black hole as described by the Cardy-Verlinde formula due to the
self-gravitation effect. These are obtained in the framework of KKW
analysis and we have kept up to linear terms in the energy of the
emitted massless particle. The afore-mentioned gravitational
background is treated as a dynamical one and the self-gravitational
corrections to its entropy are found to be positive which is a direct
violation to the holographic bound. These results are valid for two distinct
horizons $r_{\pm}$. The physical consequences are changed in the case of a single horizon and the case of no horizon, as follows. 

If $r_+=r_-$ (single horizon) then all the results discussed in
the paper hold provided $f(r_{min})=0$ \cite{akbar}.
In the case of a naked BTZ black hole (no horizon), the above results do not hold because in order the above results be valid the presence of at least one horizon is mandatory.

It is true that the corrections to Cardy-Verlinde formula have been
obtained for different types of black holes by numerous techniques
including generalized uncertainty principle (GUP) \cite{GUP},
space non-commutativity (SNC) \cite{SNC}, corrections due to thermal and quantum effects \cite{TQE} and self-gravitational
ways \cite{SGW}. The GUP corrections involve variation in the
black hole temperature with a prefactor $\alpha^2$ (where $\alpha$ is of
order unity) which also contributes negligible. Similarly the SNC
corrections barely contribute to the CFT entropy since the
non-commutativity parameter is much smaller than the conventional length
scale of black hole like its radius. The self gravitational
corrections however, as presented here, provide new insights about the CFT entropy i.e. the holography bound is not universal in the framework of two dimensional gravity models.

\section*{Acknowledgments}
This work  has been supported by the Research office of Azarbaijan
University of Tarbiat Moallem, Tabriz, Iran. 


\end{document}